\documentclass{aa}
\usepackage{graphicx}
\usepackage{txfonts}
\usepackage{natbib}

\newcommand{\chph}{Chem.~Phys.}        
\newcommand{\jpc}{J.~Phys.~Chem.}      
\newcommand{\fdis}{Faraday~Discuss.}   
\begin{document}
\title{Formation of simple organic molecules in inner T Tauri disks}
\titlerunning{Simple organics in inner T Tauri disks}
\authorrunning{Ag\'undez, Cernicharo \& Goicoechea}

\author{Marcelino Ag\'undez\inst{1}, Jos\'e Cernicharo\inst{1}, and Javier R. Goicoechea\inst{2}}

\offprints{M. Ag\'undez}

\institute{Departamento de Astrof\'isica Molecular e Infrarroja,
Instituto de Estructura de la Materia, CSIC, Serrano 121, 28006
Madrid, Spain; \email{marce@damir.iem.csic.es,
cerni@damir.iem.csic.es} \and LERMA-LRA, UMR 8112, CNRS,
Observatoire de Paris and \'Ecole Normale Sup\'erieure, 24 rue
Lhomond, 75231 Paris Cedex 05, France; \email{javier@lra.ens.fr}}

\date{Received ; accepted }


\abstract
{}
{We present time dependent chemical models for a dense and warm
O-rich gas exposed to a strong far ultraviolet field aiming at
exploring the formation of simple organic molecules in the inner
regions of protoplanetary disks around T Tauri stars.}
{An up-to-date chemical network is used to compute the evolution
of molecular abundances. Reactions of H$_2$ with small organic
radicals such as C$_2$ and C$_2$H, which are not included in
current astrochemical databases, overcome their moderate
activation energies at warm temperatures and become very important
for the gas phase synthesis of C-bearing molecules.}
{The photodissociation of CO and release of C triggers the
formation of simple organic species such as C$_2$H$_2$, HCN, and
CH$_4$. In timescales between 1 and 10$^4$ years, depending on the
density and FUV field, a steady state is reached in the model in
which molecules are continuously photodissociated but also formed,
mainly through gas phase chemical reactions involving H$_2$.}
{The application of the model to the upper layers of inner
protoplanetary disks predicts large gas phase abundances of
C$_2$H$_2$ and HCN. The implied vertical column densities are as
large as several 10$^{16}$ cm$^{-2}$ in the very inner disk ($<$ 1
AU), in good agreement with the recent infrared observations of
warm C$_2$H$_2$ and HCN in the inner regions of IRS 46 and GV Tau
disks. We also compare our results with previous chemical models
studying the photoprocessing in the outer disk regions, and find
that the gas phase chemical composition in the upper layers of the
inner terrestrial zone (a few AU) is predicted to be substantially
different from that in the upper layers of the outer disk ($>$ 50
AU).}

\keywords{astrochemistry -- stars: circumstellar matter --
planetary systems: protoplanetary disks -- ISM: molecules}

\maketitle
%

\section{Introduction}

Protoplanetary disks (PPDs) represent an intermediate stage in the
evolution from dark clouds towards planetary systems. A detailed
characterization of the physical and chemical conditions in such
disks is of great interest as they provide the initial conditions
for planet formation (see \citealt{naj07,ber07} for recent reviews
on the subject). Information about the disk chemical composition
has mainly come from radio observations, which are sensitive to
the outer cool disk, from $\sim$ 100 AU up to $\sim $900 AU
\citep{dut97,thi04}. Only recently the inner regions of PPDs have
been probed by means of infrared observations revealing the
presence of warm C$_2$H$_2$ and HCN with large gas phase
abundances \citep{lah06,gib07} and of complex organic species such
as PAHs \citep{gee06,hab06}.

Chemical models have mostly concentrated in the study of the outer
disk regions (e.g. \citealt{aik99}), in part motivated by radio
observations of several molecular species. Only a few models have
focused on the inner disk ($<$ 10 AU) chemistry. \citet{wil98}
studied the chemistry in the disk midplane, shielded from stellar
and interstellar UV photons, and found that some organic species
could be easily formed on grain surfaces. \citet{mar02} studied
the chemistry in the inner 10 AU including stellar and
interstellar UV radiation, stellar X-rays and radionuclides decay
as sources of ionization, and adsorption/desorption on grains but
not mantle chemistry, i.e. formation and destruction of molecules
occurs only in the gas phase. They calculated large abundances of
organic species such as CH$_4$ but they did not discuss the main
chemical routes to form them. \citet{sem04} and \citet{ilg06}
extensively discussed the ionization degree over a wide range of
radii and heights over the midplane. \citet{woo07} predicted that
benzene formation can be efficient in the midplane of the inner
disk ($<$ 3 AU), a dense region protected against UV photons where
grain surface reactions play an important role in building-up a
complex organic chemistry.

In this Paper we investigate gas phase routes to form simple
organic molecules, such as acetylene (C$_2$H$_2$), hydrogen
cyanide (HCN), and methane (CH$_4$), in a dense and warm
oxygen-rich gas exposed to a strong far-UV (FUV) field (h$\nu$ $<$
13.6 eV). We apply this model to the inner region ($<$ 10 AU) of a
protoplanetary disk around a T Tauri star and compare with the
abundances recently derived from observations.

\section{Chemical model}

Physical models of protoplanetary disks suggests a flared-up
structure where both the gas density and temperature vary
enormously depending on the radius $r$ from the star and height
$z$ over the disk equatorial plane (see \citealt{dul07} for a
review). The gas density decreases radially outward as a power law
of $r$, and in the vertical direction it decreases as $z$
increases in roughly an exponential way. The gas heating across
the disk is dominated by the stellar irradiation of the surface
layers at large radii ($r$ $\geq$ 10 AU), thus the kinetic
temperature increases with $z$, while at small radii ($r$ $\sim$ a
few AU) viscous dissipation may become an important heating
mechanism in the midplane regions \citep{dal98,dal99}. A flared
disk is significantly exposed to the strong UV field from the
central star. The extinction of stellar UV radiation is large in
the midplane, but it rapidly decreases as $z$ increases.
Therefore, there exists a layer of low $A_V$ where the disk
material is being photoprocessed \citep{aik99,wil00}. Here we
focus on the chemistry that takes place in the photon-dominated
region (PDR) of the inner disk ($r$ $<$ 10 AU), where typically
temperatures are several hundreds of K and gas densities range
from 10$^6$ to 10$^{11}$ cm$^{-3}$.

In order to qualitatively understand the chemical routes to form
simple organic molecules in the PDR of the inner disk, we have
firstly performed a few time dependent chemical models in which
chemical abundances evolve under fixed physical conditions
representative of this region. We consider an O-rich gas with all
the hydrogen initially as H$_2$, all the carbon as CO, the oxygen
in excess as H$_2$O and all the nitrogen as N$_2$. We use solar
abundances \citep{asp05} and assume that 50 \% of C and 65 \% of O
are in carbon and silicate grains respectively. Therefore, the
initial abundances relative to the total number of H nuclei
$n_{\rm H}$, where $n_{\rm H}$ = 2 $n$(H$_2$) + $n$(H), are
$x$(CO) = 1.25 $\times$ 10$^{-4}$, $x$(H$_2$O) = 3.5 $\times$
10$^{-5}$, $x$(N$_2$) = 3.0 $\times$ 10$^{-5}$, and $x$(He) = 8.5
$\times$ 10$^{-2}$. We adopt a gas density of $n_{\rm H}$ = 2
$\times$ 10$^8$ cm$^{-3}$ and run several models with five
different kinetic temperatures: $T_k$ = 100, 300, 500, 750 and
1000 K.

We first consider a \emph{FUV illuminated} model with a FUV field
strength of $\chi$ = 50,000 (relative to the Draine interstellar
radiation field; \citealt{dra78}), the value at 10 AU according to
FUV observations of various T Tauri disks \citep{ber03,ber04}. For
this FUV field and gas density, photoprocessing of the gas occurs
in a range of visual extinctions $A_V$ $\sim$ 0.1 - 5. If $A_V$
$\lesssim$ 0.1 the gas is very exposed to the UV radiation and
molecules are destroyed, and in the case of $A_V$ $\gtrsim$ 5
photoprocessing occurs only marginally. Here we investigate the
chemical evolution at a mean extinction value of $A_V$ = 2.5. We
then consider a separate \emph{X-ray illuminated} model, in which
the gas is solely exposed to X-rays but not FUV photons, aiming at
evaluating the influence of these two energy sources on the
chemistry. There is evidence for X-ray emission in PPDs
\citep{fei99}. Its main effect on the chemistry is the ionization
of the gas producing high energy photoelectrons (ph-e$^-$) which
further ionize and dissociate the species. To simulate this effect
we have enhanced the cosmic rays ionization rate by a factor of
1000 over the standard interstellar value, thus we take $\zeta$ =
1.2$\times$10$^{-14}$ s$^{-1}$, which is reasonable for the inner
regions of PPDs \citep{ige99}.

\begin{table}
\caption{Species included in the model} \label{table-molecules}
\centering
\begin{tabular}{lllllll}
\hline \hline
H      & C$_2$H     & NH$_2$  & C+      & CH$_5$+     & HCO+     & CN+      \\
H$_2$  & C$_2$H$_2$ & NH$_3$  & C-      & C$_2$+      & HOC+     & CN-      \\
He     & O$_2$      & CN      & O+      & C$_2$H+     & H$_2$CO+ & HCN+     \\
C      & OH         & HCN     & O-      & C$_2$H$_2$+ & H$_3$CO+ & HCNH+    \\
O      & H$_2$O     & HNC     & N+      & C$_2$H$_3$+ & CO$_2$+  & N$_2$H+  \\
N      & CO         & H$_2$CN & H$_2$+  & OH+         & HCO$_2$+ & H$_2$NC+ \\
CH     & CO$_2$     & NO      & H$_3$+  & OH-         & N$_2$+   & CNC+     \\
CH$_2$ & HCO        & e-      & CH+     & H$_2$O+     & NH+      & NO+      \\
CH$_3$ & H$_2$CO    & H+      & CH$_2$+ & H$_3$O+     & NH$_2$+  & HNO+     \\
CH$_4$ & N$_2$      & H-      & CH$_3$+ & O$_2$+      & NH$_3$+  & HNC+     \\
C$_2$  & NH         & He+     & CH$_4$+ & CO+         & NH$_4$+  &          \\
\hline
\end{tabular}
\end{table}

\begin{figure}
\begin{center}
\includegraphics[angle=0,scale=.52]{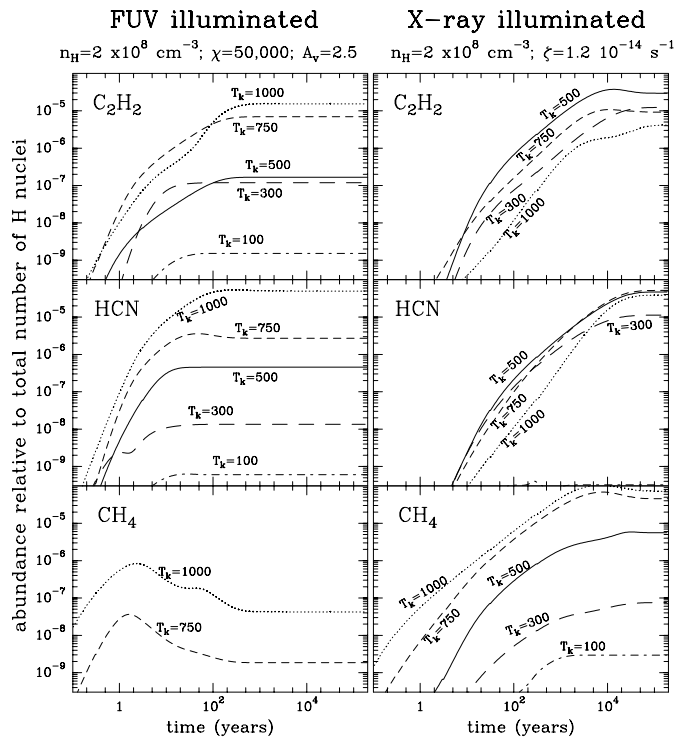}
\caption{Evolution of C$_2$H$_2$, HCN and CH$_4$ abundances for a
chemistry driven by FUV photons (left) and by X-rays (right).
Different curves correspond to different gas temperatures.}
\label{fig-abun}
\end{center}
\end{figure}

The chemical code used has been described in \citet{cer04}. The
species included in the model are given in
Table~\ref{table-molecules}. They were selected on the basis of
previous calculations including a larger number of species which
allowed us to identify the most important chemical processes
forming C-bearing species. The chemical network used was taken
from \citet{cer04} and \citet{agu06}, and was checked against the
NIST Chemical Kinetics Database\footnote{See
http://kinetics.nist.gov/kinetics/index.jsp} and the latest
version of the UMIST database for astrochemistry\footnote{See
http://www.udfa.net} \citep{woodall07}. We include neutral-neutral
and ion-molecule bimolecular reactions, three body processes and
thermal dissociations.

Important reactions involved in the built-up of small organic
molecules at high temperatures are those of H$_2$ with radicals
such as C$_2$ and C$_2$H. These reactions have been studied in the
laboratory over a relatively wide temperature range: C$_2$ + H$_2$
$\rightarrow$ C$_2$H + H between 295 and 493 K \citep{pit82} and
C$_2$H + H$_2$ $\rightarrow$ C$_2$H$_2$ + H in the range 178-440 K
\citep{pee96,opa96}. They have moderate activation barriers, of
about 1400 K, which make them very slow in the cold interstellar
medium, although at high temperatures they become rapid enough to
control the abundance of C-bearing species. Their rate constant
expressions are not included in current astrochemical databases,
which are somewhat biased toward low temperature chemistry, but
can be found in Table 1 of \citet{cer04}. The formation of H$_2$
on grain surfaces is included with a rate constant of 3 $\times$
10$^{-17}$ cm$^3$ s$^{-1}$.

\begin{figure*}
\begin{center}
\includegraphics[angle=0,scale=.62]{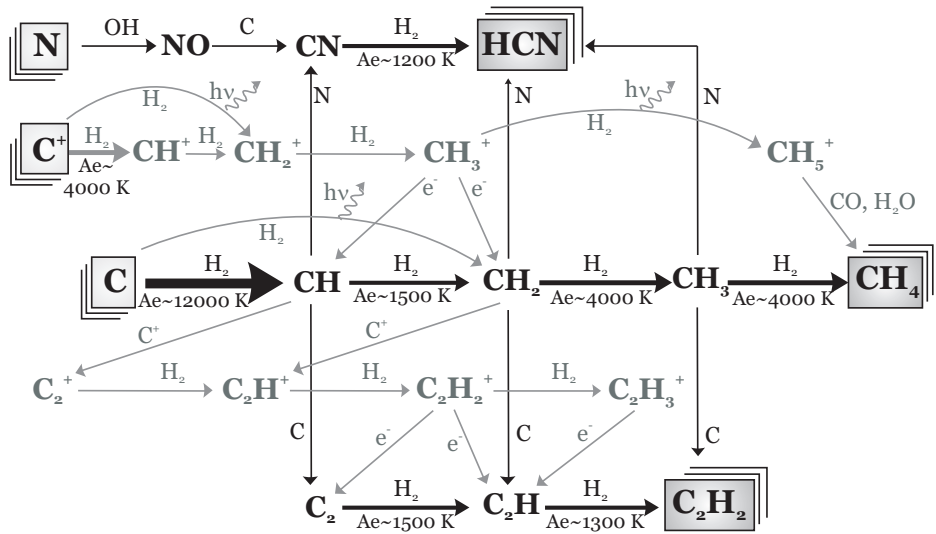}
\caption{Scheme with the main synthetic routes for the formation
of C$_2$H$_2$, HCN and CH$_4$ from C, C$^+$ and N. Reactions with
a high activation energy (Ae) are indicated by a thick arrow.}
\label{fig-reac-scheme}
\end{center}
\end{figure*}

In order to properly take into account FUV line self-shielding, we
explicitly calculated a grid of H$_2$ and CO photodissociation
rates and C photoionization rates, appropriate for the physical
conditions in the inner disk. This calculation was performed with
the \textit{Meudon PDR code} \citep{bou93}. The model has been
described in detail elsewhere \citep{pet06,goi07}. In particular,
we computed the depth dependent FUV radiation field for the range
of parameters ($\chi$, $n_H$, $T_k$ and initial abundances)
considered in the \emph{FUV illuminated} model. H$_2$, CO and C
photorates were then consistently integrated over these fields and
used afterwards in the time-dependent chemical calculations. Dust
properties are known to influence the resulting rates (e.g.
\citealt{goi07}). Here we have assumed the standard grain size
distribution proposed for the ISM \citep{mat77}. All the other
photorates included in the \emph{FUV illuminated} model have been
taken from the UMIST 2006 database. In the \emph{X-ray
illuminated} model we have included reactions induced by X-rays,
for which we have adopted the cosmic ray induced reactions rates
enhanced by a factor of 1000.

\section{Chemical routes to C$_2$H$_2$, HCN and CH$_4$}

Left panels in Fig.~\ref{fig-abun} show the evolution of
C$_2$H$_2$, HCN, and CH$_4$ abundances for several gas
temperatures in the \emph{FUV} model. The synthesis of these
organic species can be divided into three steps.

(i) Photodissociation of CO and N$_2$ with release of atomic C,
C$^+$ (produced by further photoionization of C) and atomic N. The
time scale of this step depends only on the FUV field strength,
thus on $\chi$ and $A_V$.

(ii) Atom$\rightarrow$molecule transition, from the primary
species C, C$^+$ and N to simple C-containing molecules. Important
reactions which drive this transition are (see
Fig.~\ref{fig-reac-scheme}) radiative associations such as C +
H$_2$ $\rightarrow$ CH$_2$ + h$\nu$ and C$^+$ + H$_2$
$\rightarrow$ CH$_2^+$ + h$\nu$, and rapid neutral-neutral
reactions such as C + NO $\rightarrow$ CN + O, where NO comes from
the reaction between N and OH. These reactions are nearly
temperature independent, thus their rate depends basically on the
gas density: the higher the gas density the faster they proceed.
Bimolecular reactions such as C$^+$ + H$_2$ $\rightarrow$ CH$^+$ +
H and C + H$_2$ $\rightarrow$ CH + H begin to overcome their
activation barriers at temperatures above 400 and 700 K
respectively. Three body reactions are not competitive, compared
to radiative associations, for gas densities below $\sim$
10$^{13}$ cm$^{-3}$.

(iii) Processing of simple molecules, CH$_2$, CH$_2^+$ and CN,
into more complex species, represented by C$_2$H$_2$, HCN and
CH$_4$ in our model. As Fig.~\ref{fig-reac-scheme} shows this
processing occurs through bimolecular reactions involving H$_2$, C
and N. Reactions of neutral species with H$_2$ have energy
barriers higher than 1000 K, which introduces a strong temperature
dependence. This is very marked for CH$_4$ because its formation
involves the reactions with the highest energy barriers, making it
abundant only at temperatures above $\sim$ 700 K. The synthesis of
organic molecules in the O-rich gas requires that atomic carbon,
produced by the dissociation of CO, incorporates into C-bearing
species faster than reverting to CO. This is achieved by different
mechanisms at low and high temperatures. Below $\sim$ 400 K atomic
oxygen is not converted into OH (the reaction O + H$_2$
$\rightarrow$ OH + H has an activation energy of $\sim$ 5000 K)
and the main CO-forming reaction, OH + C $\rightarrow$ CO + H, is
inhibited. Thus atomic carbon can react with other species instead
of reverting to CO. The main obstacle to form C-bearing species at
low temperature is the energy barrier of several reactions. For
example, C$_2$H$_2$ and HCN reach low abundances at 100 K because
the reactions of C$_2$H and CN with H$_2$ have activation
barriers. Above $\sim$ 400 K atomic oxygen is efficiently
converted into OH, which may react with C to form CO but reacts
faster with H$_2$ to form water. Thus, most of the oxygen forms
H$_2$O, and CO does not reach its maximum abundance allowing
atomic carbon to form C-bearing molecules. At high temperatures
most of the reactions shown in Fig.~\ref{fig-reac-scheme} overcome
their energy barriers producing a rich C-based chemistry.

The chemistry driven by X-rays (see right panels in
Fig.~\ref{fig-abun}) is rather similar to that initiated by FUV
photons except that the dissociation of CO and N$_2$, step i, is
carried out by collisions with photoelectrons: CO + ph-$e^-$
$\rightarrow$ C + O and N$_2$ + ph-e$^-$ $\rightarrow$ N + N and
by reactions with He$^+$: CO + He$^+$ $\rightarrow$ C$^+$ + O + He
and N$_2$ + He$^+$ $\rightarrow$ N + N$^+$ + He. For our chosen
FUV and X-ray field strengths, these reactions are about 100 times
slower than photodissociations, which makes the chemical time
scale to be increased by roughly the same factor, i.e. molecules
form later (see Fig.~\ref{fig-abun}). On the other hand, the lower
dissociating strength of X-rays, compared to that of the FUV
field, results in a higher steady state abundance of C-bearing
molecules. This occurs because these steady state abundances
depend on the balance between the rates at which molecules are
formed (through chemical reactions) and are destroyed (by either
FUV photons or X-rays). The destruction rates are lower in the
\emph{X-ray} model, compared to the \emph{FUV} model, and thus
chemical reactions have more time to form molecules until steady
state is reached. As a consequence, in the \emph{X-ray} model
C$_2$H$_2$ and HCN reach high abundances with little dependence on
the temperature, once this is above a threshold value of about 300
K (necessary for some important reactions to overcome their energy
barriers). Thus, the main differences in the \emph{FUV} and
\emph{X-ray} models arise because of the different dissociating
strengths. In fact, if the ionization rate is enhanced by a factor
of 100 in the \emph{X-ray} model, then the evolution of the
molecular abundances and their steady state values become much
closer to those of the \emph{FUV} model.

\begin{figure*}
\begin{center}
\includegraphics[angle=-90,scale=.72]{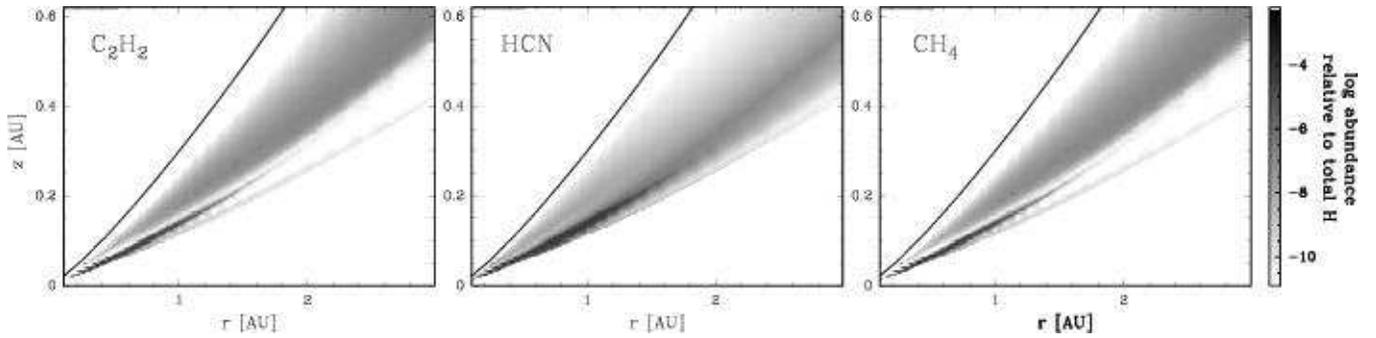}
\caption{Distribution of C$_2$H$_2$, HCN and CH$_4$ abundances in
the photodissociation region of the inner 3 AU of a protoplanetary
disk. The solid line indicates the disk surface, as given by the
height z$_{\infty}$ (see \citealt{dal99}).} \label{fig-abun-2d}
\end{center}
\end{figure*}

The relative intensity between the FUV and X-ray fields will
establish whether the chemistry is dominated by FUV photons (PDR)
or X-rays (XDR). X-rays will drive the chemistry in those regions
of protoplanetary disks with a column density toward the star
$\gtrsim$ 10$^{22}$ cm$^{-2}$, which are shielded against FUV
photons ($A_v$ $>$ 5) but not against X-rays \citep{mal96}. In
regions with column density values $\gtrsim$ 10$^{25}$ cm$^{-2}$
X-rays are severely attenuated and do not affect to the chemistry.

In summary, according to the models the formation of acetylene,
hydrogen cyanide and methane occurs in time scales between a few
years and a few thousands of years, and is clearly favored at high
temperatures. They reach a steady state in which they are
destroyed by photodissociation (or X-ray induced reactions) and
reactions with H but are continuously re-formed through reactions
involving H$_2$. Thus, the relative rates between re-formation and
destruction will determine the steady state abundance. A high
abundance of H$_2$, maintained due to its continuous formation on
grain surfaces, is essential for accelerating the process
\begin{equation}
A + H_2 \rightarrow AH + H
\end{equation}
\noindent where A is an atom or simple radical. In this way
organic species AH can reach a high steady state abundance. In
fact, if the formation of molecular hydrogen on grain surfaces is
suppressed, then H$_2$ is effectively dissociated (due to its
continuous participation in reactions of the above type) in about
10$^3$-10$^4$ years in the \emph{FUV} model and 10$^6$ years in
the \emph{X-ray} model, a time at which most of the molecules are
also destroyed, thus becoming transient species.

The steady state abundances of C$_2$H$_2$, HCN and CH$_4$ are very
sensitive to the kinetic temperature (especially in the \emph{FUV}
model), and may vary from $<$10$^{-10}$ at 100 K up to several
10$^{-5}$ at 1000 K. The chemistry of these simple organic
molecules also depends, although to a lesser extent, on other
parameters of the model. The FUV/X-ray field and gas density
affect the chemical time scale (a decrease in any of these
parameters will make the chemistry proceed slower) and also to the
steady state abundance (higher $n_{\rm H}$/$\chi$ or $n_{\rm
H}$/$\zeta$ ratios favor the formation of molecules compared to
the destruction by FUV photons or X-rays and thus favor a higher
steady state abundance). The C/O ratio is also an important
parameter which affects the wealth of C-bearing molecules.
Obviously, values close to 1 favor a rich C-based chemistry while
low values make it more difficult. We nevertheless find that only
adopting C/O ratios substantially lower than the solar value
(0.54; \citealt{asp05}), then the abundances of C-bearing species
turn to decrease appreciably. The choice of the initial abundances
is also an issue when modelling the chemistry in protoplanetary
disks and several options have been used in the literature (see
e.g. \citealt{wil98,sem04}). If we assume an initial atomic
composition, instead of a molecular one, we find that the chemical
timescale is noticeably reduced (the step i has been already
carried out) although the steady state abundance is essentially
the same. The non-dependence of steady state abundances on the
initial composition is a consequence of the strong photoprocessing
of the material, and indicates that molecular abundances in the
PDR of inner disks are not affected by the chemical history of the
gas. The actual situation in PPDs may be different if mixing
motions bring material from outer cold regions and affect the
chemical balance in the PDR.

At this point it would be interesting to discuss our results by
comparing with some other PDR models appeared in the literature. A
detailed study of the chemistry in PDRs has been carried out by
\citet{ste95}. They modelled a dense ($n_{\rm H}$ = 10$^6$
cm$^{-3}$) and highly irradiated ($\chi$ = 2 $\times$ 10$^5$)
plane parallel cloud. They considered a reduced set of chemical
species, including S- and Si-bearing molecules, of low chemical
complexity (e.g., they did not included species with more than one
carbon atom). The calculated steady state abundances for C-bearing
molecules such as HCN and CH$_4$ are lower than 10$^{-9}$ in the
warm region of $A_V$ = 0-2 ($T_k$ = 3000-50 K), although radicals
such as CH have abundances as high as 10$^{-6}$. The most
remarkable difference between their model and ours lies in the
$n_{\rm H}$/$\chi$ ratio. The much lower value of their model (5
compared to the value of 4000 adopted by us) favors photoprocesses
compared to chemical reactions and results in low abundances of
closed-shell organic molecules. A more recent PDR model by
\citet{tey04} and \citet{pet05}, focuses on the study of long
carbon chains in PDRs such as the Horsehead Nebula. They consider
a model with a gas density $n_{\rm H}$ = 2 $\times $10$^4$
cm$^{-3}$ and a FUV field strength $\chi$ = 60, thus the $n_{\rm
H}$/$\chi$ ratio is 300, much closer to the value adopted by us.
Their model nevertheless predicts C-chains such as C$_2$H and
C$_4$H to have abundances $\leq$ 10$^{-8}$. Besides a different
chemical network used, another important difference between their
models and ours resides in the kinetic temperature. In their model
the abundance of hydrocarbon radicals peaks in the region of $A_V$
= 1-2, where $T_k$ is lower than 100 K. At these relatively low
temperatures hydrocarbons do not reach high abundances since their
formation is inhibited by various reactions which have moderate
activation barriers.

\section{Abundances in the inner region of a T Tauri disk}

The physical conditions adopted in the chemical models discussed
in the previous section are assumed to be representative of those
prevailing in the PDR of inner T Tauri disks. Nevertheless, real
conditions at different points ($r$, $z$) in such disks span over
a wide range around our adopted values. In order to check whether
photochemistry can account for the formation of simple organic
molecules with abundances comparable to those observed, we have
adopted the disk physical structure, gas density and temperature
at each point, from a steady state flared accretion disk model
provided by P. D'Alessio \citep{dal98,dal99}. The disk model has a
mass accretion rate $\dot{M}$ = 10$^{-8}$ $M_{\odot}$ yr$^{-1}$,
grain properties similar to those of interstellar dust (maximum
dust grain size is 0.25 $\mu$m), and a central star with $M_*$ =
0.7 $M_{\odot}$, $T_*$ = 4000 K and $R_*$ = 2.6 $R_{\odot}$. The
chemical model was run for a grid of 18 $\times$ 16 ($r$ $\times$
$z$) points covering the upper layers of the very inner disk $r$ =
0.1-3 AU. The modelled region covers the CO/C/C$^+$ transition,
from an inner height $z_{in}$ (where the gas is well shielded
against stellar and interstellar light and all the carbon is as
CO) up to the disk surface defined by the height z$_{\infty}$
(where the gas is completely exposed to the FUV field and all the
carbon is as C$^+$, see Fig.~\ref{fig-abun-2d}).

We consider that the gas is affected by stellar and interstellar
FUV radiation and by cosmic rays. For simplicity we neglect the
effect of stellar X-rays, although in the case of strong X-ray
emitter T Tauri stars the abundances of certain molecules, which
are sensitive to the ionization rate such as HCO$^+$ or HNC, may
be substantially affected. The influence of an enhanced ionization
rate on the molecular abundances has been discussed by
\citet{aik99} and \citet{mar02}. Concerning the formation of
simple organics such as C$_2$H$_2$, HCN and CH$_4$, an enhanced
ionization rate does not greatly change the results from those
obtained when only a FUV radiation field is considered.

We treat the FUV field considering that each point ($r$, $z$) of
the disk is exposed to two separate radiation fields, stellar and
interstellar. The FUV flux emanating from the star is diluted
geometrically and attenuated by the column density of material in
the line of sight toward the point ($r$, $z$). Interstellar
radiation is assumed to penetrate into the disk in the vertical
direction. For simplicity the radiative transfer of stellar and
interstellar FUV photons is treated separately, in a plane
parallel 1+1D approach. The computed depth-dependent H$_2$, CO and
C photorates are then used in the grid of time-dependent chemical
models. Although a coupling between the attenuation of the FUV
field and  the molecular abundances exists, a self-consistent 2D
solution of the radiative transfer problem, together with the
chemical evolution, is beyond the scope of this study.

\begin{figure*}
\begin{center}
\includegraphics[angle=-90,scale=.69]{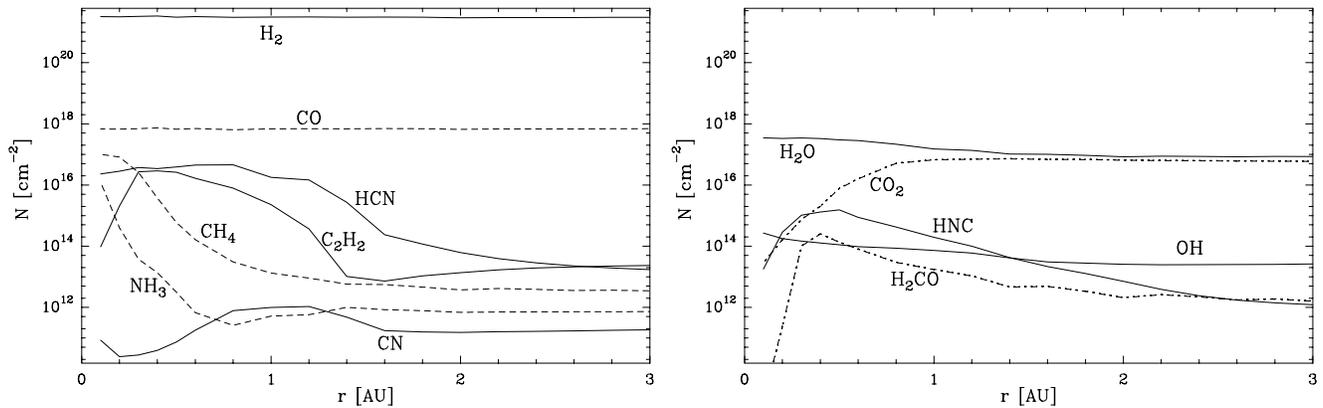}
\caption{Vertical column densities within the PDR of the inner 3
AU of a protoplanetary disk. They are computed by integrating the
gas molecular densities from the inner height $z_{in}$ to the disk
surface z$_{\infty}$.} \label{fig-cdppd-nox}
\end{center}
\end{figure*}

The spectral shape of the FUV field emitted by a T Tauri star
significantly differs from that of the interstellar radiation
field (ISRF). Therefore, photodissociation/ionization rates can be
drastically different in the presence of a T Tauri radiation field
compared to an enhanced ISRF. This problem has been addressed by
\citet{van06}, who calculated photodissociation/ionization rates
of various species for a 4000 K blackbody radiation field, typical
of T Tauri stars, and compared with the values obtained under the
ISRF. They found that, for the same integrated flux between 912
and 2050 \AA, the photorates of species such as H$_2$, CO or N$_2$
decrease by several orders of magnitude when the stellar radiation
field, instead of the ISRF, is used.

Given the impact of the true stellar radiation field on the
photochemistry, we have used the \textit{Meudon PDR code} to
generate a new grid of depth-dependent H$_2$ and CO
photodissociation rates and of C photoionization rates. Instead of
adopting a given ISRF enhancement, the FUV field is generated by a
T Tauri star simulated by a blackbody at 4000 K and a stellar
radius of 2.6 $R_{\odot}$. The intensity in the Ly$\alpha$ line,
at 1216 \AA, can be very important in T Tauri stars \citep{ber03},
and can have a large impact on the photorates of some species,
such as OH and CH$_4$, although it does not affect to species such
as H$_2$, CO and N$_2$, because their main photoabsorption bands
lie short of 1216 \AA. For simplicity, here we do not consider
enhanced emission at the Ly$\alpha$ line. The resulting radiation
field at the disk surface is then determined by the geometrical
dilution from the star. We obtain stellar photorates for H$_2$, CO
and C that are about 5 orders of magnitude smaller than the ISRF
photorates (for the same integrated flux in the FUV), in agreement
with the values reported by \citet{van06}. Stellar photorates for
species other than H$_2$, CO and C are taken from \citet{van06}
when available, or are assumed to be equal to the ISRF photorates
otherwise. The visual extinction of stellar and interstellar light
at a given point is assumed to be proportional to the column
density of hydrogen nuclei $N_{\rm H}$ in the direction toward the
star and outward in the vertical direction respectively. We use
the standard relation $A_V$ = $N_{\rm H}$ (cm$^{-2}$) / 1.87
$\times$ 10$^{21}$ \citep{boh78}, adequate for the ISM grain size
distribution. In the modelled region, the stellar FUV field has a
strength which is higher than the ISRF by orders of magnitude
($\chi$ = 10$^4$-10$^7$). However, since the stellar photorates
are for many species lower than the ISRF photorates by orders of
magnitude, the photoprocessing of the material in the PDR layers
is to a great extent dominated by the interstellar, rather than
stellar, FUV radiation field.

The gas densities are 10$^8$-10$^{11}$ cm$^{-3}$ in the sampled
region and the gas kinetic temperature, assumed equal to the dust
temperature in the disk physical model, ranges from 100 to 1000 K.
According to \citet{kam04}, the gas temperature significantly
exceeds the dust temperature in the very upper layers with a
visual extinction of stellar radiation $A_V$ $<$ 0.1. In this
superheated surface layer our model most probably underestimates
the gas temperature. The implications however for the calculated
abundances of simple organic species are little since this is a
region where CO has fully dissociated and most of the carbon is as
C$^+$. We do not consider adsorption/desorption on dust grains
since at the high temperatures prevailing in the studied layers
($T_k$ $>$ 100 K) all the molecules are supposed to have been
evaporated from grain mantles and to be in the gas phase.

The abundance distributions at time 10$^4$ yr, when steady state
has been reached, for C$_2$H$_2$, HCN, and CH$_4$ are plotted in
Fig.~\ref{fig-abun-2d}. A thin shell of the disk contains simple
organic species with abundances ranging from 10$^{-8}$ (in the
outer part, at 2-3 AU) up to 10$^{-4}$ (in the inner 1 AU where
gas is hot). The amount of material contained in such a thin layer
is however remarkably large due to the large gas densities
prevailing there. Vertical column densities within the disk PDR
can be calculated by integrating the gas molecular densities in
the $z$ direction (see Fig.~\ref{fig-cdppd-nox}). The column
densities of H$_2$ and CO are also shown to visualize the
molecular abundances relative to these two species.

\citet{lah06} observed toward IRS 46 hot ($>$ 350 K) C$_2$H$_2$,
HCN and CO$_2$ with column densities of 3, 5 and 10 $\times$
10$^{16}$ cm$^{-2}$ respectively. In GV Tau, \citet{gib07}
detected warm C$_2$H$_2$ and HCN with column densities of 7 and 4
$\times$ 10$^{16}$ cm$^{-2}$ respectively, and derived an upper
limit for methane of CH$_4$/CO $<$ 0.0037, rather similar to that
found by \citet{gib04} in HL Tauri (CH$_4$/CO $<$ 0.005). The
inferred warm temperatures led these authors to suggest that both
acetylene and hydrogen cyanide are most likely located in the
inner disk, within a few AU around the star. In our model, the
vertical column densities of C$_2$H$_2$ and HCN are a few
10$^{16}$ cm$^{-2}$ in the inner 1 AU but noticeably decrease with
$r$ following the decrease in temperature. The column densities of
these two species could be enhanced at radii larger than 1 AU if
the gas temperatures are higher than in our disk model or if
vertical and radial mixing brings warm material toward the cooler
midplane regions, as has been suggested to occur for CO in the
outer regions of PPDs \citep{sem06}. Our model predicts CH$_4$ to
have a large abundance in the very inner disk ($<$ 0.5 AU).
However, it is less abundant than C$_2$H$_2$ and HCN throughout
the rest of the disk, unlike in the model by \citet{mar02} where
it is the most abundant organic species. The reasons of such
difference are unclear. In the model by \citet{mar02} the large
gas phase abundance of methane is likely due to its desorption
from grains surfaces. However, since they do not include formation
of molecules on grain surfaces, CH$_4$ must have been previously
formed through a mechanism of gas phase reactions which is not
explicitly detailed. The regions modelled by them and us are
however different as we only consider the upper disk layers
exposed to FUV radiation while they consider the full disk in the
$z$ direction down to the midplane, where most of material is
contained. Thus, the vertical column densities of H$_2$ in our
modelled region are $\sim$ 10$^{21}$ cm$^{-2}$ while in their
model are $\sim$ 10$^{25}$ cm$^{-2}$, so that the large abundances
calculated by them are coming from a region much closer to the
midplane than that studied by us, which makes difficult a direct
comparison between both models.

In our model, CO$_2$ is formed very efficiently through the
reaction between CO and OH and has a fairly large column density,
about 10$^{17}$ cm$^{-2}$, which agrees with the value observed
toward IRS 46 by \citet{lah06}. Fig.~\ref{fig-cdppd-nox} also
gives the column densities of some other molecules which, although
have not been observed in the inner regions of T Tauri disks, are
predicted to have relatively large column densities in the PDR.
H$_2$CO is formed with a moderate abundance through various
reactions involving atomic oxygen and small hydrocarbon radicals.
The presence of HNC is strongly related to that of HCN, since it
is formed through proton transfer to the latter species and
subsequent dissociative recombination. The column density of
NH$_3$ has a radial profile similar to that of CH$_4$, but with a
value about one order of magnitude lower. Also it is interesting
to note that the abundance ratio of a closed-shell molecule and
its related radical (e.g. H$_2$O/OH and HCN/CN) is quite large
(about 100-1000), in contrast with the values close to 1 predicted
in the outer disk, $r$ $>$ 50 AU, (e.g. \citealt{wil00}). This is
a consequence of the large gas densities prevailing in the inner
disk which, in spite of the strong FUV field, tend to favor the
formation of closed-shell species through reaction~(1).

\section{Conclusions}

In summary, we have shown that a strong FUV/X-ray field may
efficiently drive a rich C-based chemistry in a dense O-rich gas
with kinetic temperatures of several hundreds of degrees Kelvin.
The application of our PDR chemical model to the inner regions of
a T Tauri disk yields gas phase abundances for simple organic
molecules such as C$_2$H$_2$ and HCN that range from 5 $\times$
10$^{-5}$ to 10$^{-8}$, depending on the radius. This translates
into vertical column densities as large as 10$^{17}$ cm$^{-2}$ in
the very inner disk ($<$ 1 AU) down to 10$^{13}$ at 3 AU. The
model thus explains the large column densities of C$_2$H$_2$ and
HCN (several 10$^{16}$ cm$^{-2}$) observed in the inner regions (a
few AU) of IRS 46 and GV Tau disks \citep{lah06,gib07}. The huge
variations with radius of the gas density and temperature within
the PDR layer of protoplanetary disks results in a quite different
gas phase chemistry, i.e. different molecular abundances and
ratios between them, in the inner terrestrial zone compared to the
outer regions at several hundreds of AU. Our results await
confirmation from further observations of different molecular
species probing the inner regions of protoplanetary disks.

\begin{acknowledgements}

We are grateful to the anonymous referee for useful comments and
suggestions which greatly helped to improve this paper. We are
indebted to P. D'Alessio for kindly providing the physical model
of the PPD. We also thank C. Ceccarelli and E. Herbst for very
useful comments on a previous version of this manuscript. This
work has been supported by Spanish MEC through grants
AYA2003-2785, AYA2006-14876 and ESP2004-665 and by Spanish CAM
under PRICIT project S-0505/ESP-0237 (ASTROCAM). MA also
acknowledges grant AP2003-4619 from Spanish MEC. JRG was supported
by a \textit{Marie Curie Intra-European Individual Fellowship}
within the 6th European Community Framework Programme, contract
MEIF-CT-2005-515340.
\end{acknowledgements}

\end{document}